\magnification=1200
\baselineskip=20pt

\line{\hfil MRI-PHY-971217}

\def\lam{\Lambda}
\def\glmunu{G^a_{\mu\nu}}
\def\gumunu{G^{a\mu\nu}}
\def\gtumunu{\tilde {G}^{a\mu\nu}}
\def\phid{\phi^+}
\def\as{\alpha_s}
\centerline{\bf Probing Hg contact interactions by $gg\rightarrow H$}
\centerline{\bf at a high energy hadron collider}
\vskip .4 true in

\centerline{\bf Uma Mahanta}
\centerline{\bf Mehta Research Institute}
\centerline{\bf Chhatnag Road, Jhusi}
\centerline{\bf Allahabad-221506, India}

\vskip 1 true in

\centerline{\bf Abstarct}

In this article we study the effect of Hg contact interactions on
H production by gluon fusion at a hadron collider. Two such d=6 operators
 have been considered. The present precision for measuring $\as$ implies
that the lower bound on the scale for the CP even operator $O_1$ must be 
around 2 TeV. Whereas the precision for measuring $d_n$ gives rise to a 
lower bound of about $10^5$ TeV on the scale for the CP odd operator $O_2$.
We show that for the above lower bound $O_1$ can significantly affect 
the parton level cross-section for the process $gg\rightarrow H$.

\vfill\eject

The excess of events over SM prediction reported by ZEUS and H1 experiments 
[1] at HERA in $e^+p\rightarrow e^+jX$ at $Q^2> 15000$ GeV$^2$ have been 
given several attractive interpretations namely (a) s-channel leptoquark 
exchange [2] (b) R-parity violating squark exchange [2] and (c) eq four
fermion contact interactions [3]. Assuming that the data does not favor an
s-channel resonance then the effective Lagrangian containing LL, LR, RL
and RR  eq contact interactions provide the most model independent 
parametrization of possible new physics effects at HERA enenrgy. Some of 
these works have even attributed the higgs $Q^2$ HERA excess to possible
eq compositeness whose effects at low energies can  be represented by
the four fermion Lagrangian.
If the eq 
contact interactions arise from TeV scale compositeness structure underlying
the light SM particles then analogous effects could also arise elsewhere as
for example through Hg contact interactions. The aim of this article is 
to study the effects of Hg contact interactions on Higgs production 
cross-section at a high energy hadron collider like LHC.

If the SM particles are composites of more elementary constituents associated 
with a compositeness scale $\Lambda$ where $\Lambda\gg v$, then its effects 
at lower energies can be parameterized in terms of non-renormalizable 
operators involving the light SM fields. A non-renormalizable operator 
$O_i$ of dimension $d_i$ ($d_i>4$) will be associated with a coefficient 
$C_i=\pm{a_i\over \Lambda^{d_i-4}}$. The dimensionless coefficient $a_i$
depends on the nature of new heavy physics associated with the scale
$\lam$. For $d_i=6$ the dimensionless coefficient 
$a_i$ is expected to lie between 1 and 4$\pi$ if the new physics
is intrinsically strong. However if the new heavy physics associated with
the scale $\lam$ is intinsically weak and the operator $O_i$ arises 
from loop level process then $a_i$ can be much smaller than 1. In this
work however we shall take a phenomenological approach and
not commit ourselves to either a strongly interacting or a weakly interacting
new physics scenario.
 For $d_i=6$ low energy measurements give rise to
lower bounds on the combination ${\lam\over \sqrt {\vert a_i\vert}}$ and
includes the effects of $a_i$. The estimates presented in this work are
therefore independent of the precise nature of new physics.
The $\pm$ sign associated with $C_i$
determines whether the new physics contribution to some physical amplitude
interferes constructively or destructively with the SM effect.

The discovery of the higgs boson is clearly one of the most important goals
of all high energy colliders. In SM, for a heavy top quark ($m_t> 100$ Gev)
and $m_h\le 500 $ Gev the dominant contribution [4]
to higgs productionn at LHC is expected to arise from gluon fusion mechanism
[5], which proceeds through a triangle loop diagram. At tevatron energy
the dominant higgs production mechanism is $q\bar {q}^{\prime}\rightarrow
WH$. In the gluon fusion process
 quarks of all flavor
contribute to the loop diagram with the dominant contribution coming from
heavy quarks ($m_t\ge {m_h\over 2}$). Being a loop induced process the 
production rate of a higgs boson by this process is highly sensitive to new 
physics effects. TeV scale compositeness effects could 
in principle make important
contributions to the cross-section for the process $gg\rightarrow H$. The 
effects of possible Tev scale compositeness on the process $gg\rightarrow H$
can be expressed by means of following $d=6$ operators involving $\phi$
and $\glmunu$ [$\glmunu =\partial_{\mu}G^a_{\nu}-\partial_{\nu}G^a_{\mu}
+g_sf_{abc}G^b_{\mu}G^c_{\nu}$]

$$O_1=(\phid\phi )\glmunu\gumunu ={(v+h)^2\over 2}\glmunu\gumunu.\eqno(1a)$$

$$O_2=(\phid\phi )\glmunu\gtumunu ={(v+h)^2\over 2}\glmunu\gtumunu.\eqno(2a)$$

Note that both $O_1$ and $O_2$ are invariant under the SM gauge group
$SU(3)_c\times SU(2)_l\times U(1)_y$. However while $O_1$ is even under
the discrete symmetries P and T, $O_2$ is odd under the same trasformations.
In the following we shall consider their effects on low energy processes
separately.

Effects of $O_1$: Besides contributing to ggh, gggh and ggggh vertices, $O_1$
alos modifies the canonical gluon K.E. term

$$L_{KE}=-{1\over 4}(1-{2a_1v^2\over \lam_1^2})\glmunu\gumunu.\eqno(2)$$

Here $\lam_1$ is the charateristic scale associated with the operator $O_1$.
Since $O_1$ and $O_2$ have different symmetry properties under P and T they
could aproiri be associated with different energy scales. Let us make a change
of scale of $A^a_{\mu}$: $A^a_{\mu}\rightarrow A^{\prime a}_{\mu}=
(1-{2a_1v^2\over \lam_1^2})^{1\over 2}A^a_{\mu}$ so that the free part of
$L_{KE}$ has the canonical normalized form in terms of new fields
namely $-{1\over 4}(\partial_{\mu} A^{\prime a}_{\nu}-\partial_{\nu}
 A^{\prime a}_{\mu}) (\partial^{\mu} A^{\prime a\nu}-\partial^{\nu}
 A^{\prime a\mu})$. However the ggg and gggg  interaction vertices
in $L_{KE}$ can also be brought to the canonical form if the QCD coupling
$g_s$ is simultaneously subjected to a multiplicative renormalization:
$g_s\rightarrow g^{\prime}_s={g_s\over (1-{2a_1v^2\over \lam_1^2})
^{1\over 2}}$. In the following discussion we shall drop the primes
and all quantities will refer to primed ones. In the context of the SM
$\alpha_s$ is an unknown parameter. Therefore it is not possible to 
determine a lower bound on $\lam_1$ by exploiting the renormalization
of $\as$ due to new physics. If a first principle  calculation of
$(\as )_{sm}$ is available on the basis of a more fundamental theory
 we could determine a lower bound on $\lam_1$ from the difference between the 
experimental and predicted value of $\as$. Nevertheless we can find
an approximate
lower bound on $\lam_1$ by demanding that the new physics contribution to
$\as$ be no greater than $2\sigma$ where $\sigma$ is the experimental
error in measuring
$\as$. Note that $(\as )_{sm}$ is the value of $\as$ in the limit
$\lam_1\rightarrow \infty$. The present global average of [6] of $\as$ from
different measurements is $\as =.118 \pm .003$ where .003 is the smallest
systematic error in an individual result namely in the lattice result.
We therefore require that $(\delta \as )_{new}\approx (\as )_{sm}[{1\over
(1-{2a_1v^2\over \lam_1^2})}-1]\approx \pm .006 $. from which it follows that
${\lam_1\over \sqrt {a_1}}\ge 1.5$ TeV (for $a_1> 0$) and  
${\lam_1\over \sqrt {\vert a_1 \vert}}\ge 1.1$ TeV (for $a_1<0$). Next we shall
consider the effect of $O_1$ on higgs production by gluon fusion at a
high energy hadron collider. In  SM the process $gg\rightarrow H$ takes
place through an intermediate triangular loop diagram. It can be described by
the effective Lagrangian [7] (which is obtained by integrating over all
quark flavors)

$$L_{sm}\approx -{1\over v}{\as (m^2_h)\over 12\pi}IHg^a_{\mu\nu}
g^{a\mu\nu}\eqno(3)$$
where $g^a_{\mu\nu}$ is the free part of the gluon field strength tensor.

$I=\sum_q I_q$,\ \  $I_q=3[2x_q+x_q(4x_q-1)f(x_q)]$,\ \  $x_q={m^2_q\over 
m^2_h}$

$f(x_q)=-2[Sin^{-1}{1\over2\sqrt {x_q}}]^2$ for $x_q>{1\over 4}$

$f(x_q)={1\over 2}(\ln{y_q^+\over y_q^-})^2-{\pi^2\over 2}+\imath \pi 
\ln{y_q^+\over y_q^-}$ for $y_q<{1\over 4}$ and 

$y_q^{\pm}={1\over 2}\pm \sqrt {{1\over 4}-x_q}$.

In the large $m_q$ limit ($x_q\gg 1$) $I_q\rightarrow 1$, whereas in the small
$m_q$ limit ($x_q\ll 1$)\ \  $I_q\rightarrow 0$. If there are N heavy quark
flavors with $m_q\ge m_h$ then $I\approx N$. For $100 GeV\le m_h\le 400 Gev$
only the top quark contribution is relevant in SM with three quark families.
The effective Lagrangian that contains the SM as well as new physics
sources for $gg\rightarrow H$ is

$$L_{eff}\approx {1\over v} ({a_1 v^2\over \lam_1^2}-{\as (m^2_h)\over 12\pi}I)
Hg^a_{\mu\nu}g^{a\mu\nu}\eqno(4).$$

Hence for $a_1<0$ ($a_1>0$) there is constructive  (destructive) interfernce 
between SM and new physics contributions. For $m_h=200$ GeV and $m_t=175$ GeV,
$x_q\approx .766$, $I(x_q)\approx 1$ and $\as (m_h^2)\approx .107$.
If ${\lam\over \sqrt {\vert a_1\vert}}\approx 2.5$ TeV we find that in the
destructive interfernce scenario the lowest order parton level cross-section
is 5.4 times its SM value. On the other hand for constructive interference
the lowest order cross-section becomes 18.7 times its SM value for the same 
values of the parameters. Hence if ${\lam_1\over \sqrt {\vert a_1\vert}}$
is as low as 2.5 TeV, either sign of $a_1$ yields a cross-section which
is significantly greater than the SM value. Keeping $m_h=200$ GeV if we
increase  ${\lam_1\over \sqrt {\vert a_1\vert}}$ to 5 TeV we find that 
${\sigma\over (\sigma )_{sm}}\approx .03$ for destructive interference. 
Whereas for constructive interference ${\sigma\over (\sigma )_{sm}}\approx
3.3$. Hence the ratio ${\sigma\over (\sigma )_{sm}}$ is a sensitive function
of the scale of new physics. On the other hand if we keep
  ${\lam_1\over \sqrt {\vert a_1\vert}}$ fixed at 5 TeV but increase $m_h$
to 400 GeV, the ratio  ${\sigma\over (\sigma )_{sm}}$ remains almost the 
same as in the case of $m_h =200$ GeV. 

Effects of $O_2$: We shall now consider the effect of $O_2$ on
 low energy measurements. $O_2$
changes the  $\theta$ parameter
of the effective QCD Lagrangian $L_{QCD}^{eff}=
L_{QCD}+{\theta\as\over 8\pi}\glmunu\gtumunu $ to $\theta +\delta\theta$
 where $\delta\theta={4\pi\over \as}{a_2v^2\over \lam_2^2}$. When the 
electro-weak theory is appended to QCD, due to $U(1)_A$ anomaly the
CP violating phase from the quark mass matrix slips into the the QCD theta
term. The result is that $\theta +\delta\theta$ gets replaced by 
$\bar {\theta}=\theta +\delta\theta + arg (det M)$. $\bar {\theta}$
gives rise to electric dipole moment of neutron. Current algebra estimates
[8] give $d_n\le 3.6\times 10^{-16}\bar {\theta}$. However
$\bar {\theta}$ is an unknown parameter in the context of the SM. So strictly
speaking one cannot determine a lower bound on $\lam_2$ from the experimental
bound on $\bar{\theta}$. However an approximate lower bound on $\lam_2$
can be derived by demanding that the new physics contribution to
$d_n$ be no greater than $2\sigma$ where $\sigma$ is the experimental error
in measuring [9] $d_n$. In this way we can arrive at the lower bound 
 ${\lam_2\over \sqrt {\vert a_2\vert}}\ge 85\times 10^3$ Tev which is too
large to be probed at any of the upcoming colliders.
 We shall therefore neglect the
effect of $O_2$ on higgs production by gluon-gluon annihilation.

In conclusion in this article we have considered the effect of Hg contact 
interaction onon the process $gg\rightarrow H$. Two such operators $O_1$
and $O_2$ were identified. The lower bound on the scale for the CP even
operator $O_1$ was found to be around 2 TeV. Whereas the lower bound on the
scale associated with the CP odd operator $O_2$ was found to be around 
$10^5$ TeV. The effect of $O_2$ on the process $gg\rightarrow H$  is therefore
negligible. However the operator $O_1$ affects the parton level
cross-section for the process $gg\rightarrow H$ significantly if
 ${\lam_1\over \sqrt {\vert a_1\vert}}$ lies around its lower bound for either
sign of $a_1$. For example if $a_1 > 0$ and $\lam_1\approx $ 5 TeV the effect
of new physics reduces the cross-section to .03 times its SM value
which would make the signal almost impossible to observe. On the other hand
 if $a_1 <0$ and $\lam_1\approx $ 2.5 TeV the effects of new physics
enhances the cross-section to almost 20 times its SM value.
 It should be noted that the lowest order cross-section for
$gg\rightarrow h$ is also modified due to QCD radiative corrections in the 
context of the SM. Therefore to identify the new physics contribution
from the difference between the observed and predicted values of the
cross-section a precise estimate of QCD indiuced radiative corrections to
$\sigma_{gg\rightarrow h}$ is essential.

\centerline{\bf References}

\item{1.} H1 Collaboration, C. Adolff et al., hep-ex/9702012; J. Breitweg
et al., hep-ex/9702015.

\item{2.} D. Choudhury and S. Raychaudhuri, hep-ph/9702392; H. Dreiner
and P. Morawitz, hep-ph/9703279; K. S. Babu et al., hep-ph/9703299;
J. L. Hewett and T.G. Rizzo, hep-ph/9703337; R. Barbieri et al.,
hep-ph/9704275.

\item{3.} V. Barger et al., hep-ph/9703311; G. Altarelli et al.,
hep-ph/9703276; W. Buchmuller and D. Wyler, hep-ph/9704317.

\item{4.} J. Gunion et al., The Higgs Hunter's guide, Addsion Wesley
(Menlo-Park, 1991).

\item{5.} H. Georgi et al., Phys. Rev. Lett. 40, 692 (1978); T.
Rizzo, Phys. Rev. D 22, 178 (1980).

\item{6.} Review of Particle Properties, Phys. Rev. D 54, 77 (1996).

\item{7.} V. Barger and R. Phillips, Collider Physics, 
Addision-Wesley (1987).

\item{8.} R. J. Crewther et al., Phys. Lett. B 88, 123 (1979); B 91, 487
(1980) E.

\item{9.} I. S. Altarev et al., JETP. Lett. 44, 460 (1986).

\end